\documentclass[12pt]{llncs}
\usepackage{graphicx, subfigure}
\usepackage{float}
\usepackage{algorithm}
\usepackage{algpseudocode}
\usepackage{theorem}
\usepackage{epsf}
\usepackage{epsfig}
\usepackage{multirow}
\usepackage{slashbox}
\usepackage{fullpage}

\newtheorem{lem}{Lemma}
\newcommand{\keywords}[1]{\par\addvspace\baselineskip
\noindent\keywordname\enspace\ignorespaces#1}
\begin{document}
\title{Optimal Deterministic Ring Exploration with Oblivious Asynchronous Robots}
\author{Anissa Lamani \and Maria Potop-Butucaru \and
S\'{e}bastien Tixeuil
}
\institute{Universit\'e Pierre et Marie Curie - Paris 6, LIP6-CNRS 7606, France}
\maketitle
\begin {abstract}
We consider the problem of exploring an anonymous unoriented ring of size $n$ by $k$ identical, oblivious, asynchronous mobile robots, that are unable to communicate, yet have the ability to sense their environment and take decisions based on their local view. Previous works in this weak scenario prove that $k$ must not divide $n$ for a deterministic solution to exist. Also, it is known that the minimum number of robots (either deterministic or probabilistic) to explore a ring of size $n$ is $4$. An upper bound of $17$ robots holds in the deterministic case while $4$ probabilistic robots are sufficient.
In this paper, we close the complexity gap in the deterministic setting, by proving that no deterministic exploration is feasible with less than five robots whenever the size of the ring is even, and that five robots are sufficient for any $n$ that is coprime with five. Our protocol completes exploration in $O(n)$ robot moves, which is also optimal.

\end {abstract}
\section{Introduction}\label{sec:Introduction}

Recent research focused on systems of autonomous mobile entities (that are hereafter referred to as \emph{robots}) that have to collaborate in order to accomplish collective tasks. Two universes have been studied: the continuous euclidean space \cite{paol08,yama99,yoa08} where the robots entities can freely move on a plane, and the discrete universe in which space is partitioned into a finite number of locations, conventionally represented by a graph, where the nodes represent the possible locations that a robot can take and the edges the possibility for a robot to move from one location to the other \cite{paola04,kowa04,marc06,das07,klas08,Ralf08,davi07,davi08,stap09}. In this paper we pursue research in the discrete universe and focus on the exploration problem when the network is an anonymous unoriented ring, using a team of autonomous mobile robots. The robots we consider are unable to communicate, however they can sense their environment and take decisions according to their local view. We assume anonymous and uniform robots (\textit{i.e} they execute the same protocol and there is no way to distinguish between them using their appearance). In addition they are oblivious, \textit{i.e} they do not remember their past actions.  
In this context, robots asynchronously operate in cycles of three phases: look, compute and move phases. In the first phase, robots observe their environment in order to get the position of all the other robots in the ring. In the second phase, they perform a local computation using the previously obtained view and decide on their target destination to which they will move in the last phase. 

\paragraph{\textbf{Related work}}\label{sec:relWork}

In the discrete model, two main problems are investigated assuming very weak asynchronous, identical, and oblivious robots: the gathering and the exploration problem. In the gathering problem, robots have to gather in one location not known in advance \textit{i.e} there exists an instant $t>0$ where all robots share the same location (one node of the ring). In the exploration problem, robots have to explore a given graph, every node of the graph must be visited by at least one robot and the protocol eventually terminates (that is, all robots are idle). 

For the problem of gathering in the discrete robot model, the aforementioned weak assumptions have been introduced in \cite{klas08}. The authors proved that the gathering problem is not feasible in some symmetric configurations and proposed a protocol based on breaking the symmetry of the system. By contrast in \cite{Ralf08}, they proposed a gathering protocol that exploits this symmetry for a large number of robots ($k>18$) closing the open problem of characterizing symmetric situations on the ring which admit a gathering. 

For the exploration problem, the fact that the robots have to stop after the exploration process implies that the robots somehow have to remember which part of the graph has been explored. Nevertheless, in this weak scenario, robots have no memory and thus are unable to remember the various steps taken before. In addition, they are unable to communicate explicitly, therefore the positions of the other robots remain the only way to distinguish different stages of the exploration process. The main complexity measure here is the minimal number of robots necessary in order to explore a given graph. It is clear that a single robot is not sufficient for the exploration in the case where it is not allowed to use labels. In \cite{davi08}, it has been shown that $\Omega(n)$ robots are necessary in order to explore trees of size $n$, however, when the maximum degree of the tree is equal to three then the exploration can be done with a sub-linear robot complexity. In the case where the graph is a ring, it has been shown in \cite{davi07} that $k$ (the number of robots) must not divide $n$ (the size of the ring) to enable a deterministic solution. This implies that for a general $n$, $\log(n)$ robots are necessary. The authors also present in \cite{davi07} a deterministic protocol using $17$ robots for every $n$ that is coprime with $17$. By contrast, \cite{stap09} presents a probabilistic exploration algorithm for a ring topology of size $n> 8$. Four probabilistic robots are proved optimal since the same paper shows that no protocol (probabilistic or deterministic) can explore a ring with three robots.

\paragraph{\textbf{Contribution}}\label{sec:Contribution}

In this paper, we close the complexity gap in the deterministic setting. In more details, we prove that there exists no deterministic protocol that can explore an even sized ring with $k\leq 4$ robots. This impossibility result is written for the ATOM model~\cite{yama99} where robots execute their look, compute and move phases in an atomic manner, and thus extend naturally in the non-atomic CORDA model. We complement the result with a deterministic protocol using five robots and performing in the fully asynchronous non-atomic CORDA model \cite{Pre01} (provided that five and $n$ are coprime). The total number of robot moves is upper bounded by $O(n)$, which is trivially optimal. 

\section{Model and Preliminaries}\label{sec:Model}

We consider a distributed system of mobile robots scattered on a ring of $n$ nodes $u_{0}$,$u_{1}$,..., $u_{(n-1)}$ such as $u_{i}$ is connected to both $u_{(i-1)}$ and $u_{(i+1)}$. The ring is assumed to be anonymous \textit{i.e} there is no way to distinguish the nodes or the edges (i.e. there is no available labeling). In addition, the ring is unoriented \textit{i.e} given two neighbors, it is impossible to determine which node is on the right or on the left of the other. 
On this ring $k$ robots collaborate to explore all the nodes of the ring. The robots are identical \textit{i.e} they cannot be distinguished using their appearance and all of them execute the same protocol. Additionally, the robots are oblivious \textit{i.e} they have no memory of their past actions. We assume the robots do not communicate in a explicit way. However, they have the ability to sense their environment and see the position of the other robots. Each robot can detect whether several robots are on the same node or not, this ability is called \textit{multiplicity detection}. Robots operate in three phase cycles: Look, Compute and Move. During the Look phase robots take a snapshot of their environment. The collected information (position of the other robots) are used in the compute phase in which robots decide to move or to stay idle. In the last phase (move phase) they may move to one of their adjacent nodes towards the target destination computed in the previous phase.   

At some time $t$, a subset of robots are activated by an abstract entity called \textit{scheduler}. The scheduler can be seen as an external entity which selects some robots for the execution. In the following we assume that the scheduler is fair \textit{i.e} each robot is activated infinitely many times. Two computational models exist: The \textit{ATOM model} \cite{yama99}, in which synchronous cycles are executed in atomic way \textit{i.e} the robots selected by the scheduler at the beginning of a cycle execute synchronously the full cycle, and the \textit{CORDA model} \cite{Pre01} in which the scheduler is allowed to interleave different phases (For instance one robot can perform a look operation while another is moving).
The model considered in our case is the CORDA model with the following constraint: the Move operation is instantaneous \textit{i.e} when a robot takes a snapshot of its environment, it sees the other robots on nodes and not on edges. Nevertheless, since the scheduler is allowed to interleave the operations, a robot can move according to an outdated view (during the computation phase, some robots have moved).

 


In the following we assume that initially every node  of the ring contains at most one robot. During the system execution a subset of robots are activated and move to other nodes. The position of all the robots at time $t$ is the system configuration at $t$. During the Look phase, the activated robots take a snapshot of their environment in order to see the position of the other robots. The snapshot result is called a view and is defined by the two following sequences:  $C^{+i}(t) =<d_{i}(t) d_{i+1}(t)...d_{i+n-1}(t)>$ and $C^{-i}(t) =<d_{i}(t) d_{i-1}(t)...d_{i-(n-1)}(t)>$ where $d_{i} (t)$ denotes the multiplicity of robots on the node $u_{i}$ at instant $t$ taking an arbitrarily orientation of the ring. $d_{i}>1,  \forall i\in [1,n]$  if and only if $u_{i}$ is occupied by at least one robot, $d_{i}=0$ otherwise. When $d_{i} (t)=0$, the node $u_{i}$ is said to be empty at instant $t$, when $d_{i}(t)=1$, we say that the node $u_{i}$ is occupied at instant $t$, otherwise we say that there is a \textit{tower} on $u_{i}$ at instant $t$. 

The view at $u_{i}$ is said to be \textit{symmetric} at instant $t$ if and only if $C^{+i}(t)=C^{-i}(t)$. Otherwise, the view of $u_{i}$ is said to be \textit{asymmetric}. When the configuration is symmetric, both edges incident to the node $u_{i}$ which is occupied by the robot taking the snapshot look identical. In this case we assume the worst scenario allowing the adversary to take the decision on the direction to be taken.

\paragraph{\textbf{Problem to be solved}}\label{sec:problem}

The problem considered here is the exploration problem, where $k$ robots have to collaborate and explore a ring of size $n$ before stopping forever. A protocol $P$ solves the exploration problem if and only if the following conditions are satisfied:
\begin{enumerate}
\item{\textbf{Safety}} Every node is visited by at least one robot.
\item {\textbf{Termination}} The algorithm eventually stops.
\end{enumerate}

\section{Impossibility result}

It has been shown~\cite{davi07} that no deterministic team of $k$ robots can explore a ring of size $n$ when $k$ divides $n$. Also, it is known~\cite{stap09} that no exploration protocol (deterministic or probabilistic) is possible when $0\leq k < 4$. We now observe that when a single robot is activated at a time, a trivial deterministic variation of the protocol of~\cite{stap09} (that uses four robots and randomization to break symmetry in some situations) matches the lower bound. So, our leveraging of the lower bound result of~\cite{stap09} to four robots considers the case when several robots can be activated at the same time.

\begin{lem}
There exists no deterministic protocol for exploring a ring of an even size ($n$) with four robots.
\end{lem}

\begin{proof}
The proof is by contradiction. We assume that there exists a deterministic protocol with four robots that can explore a ring and terminate. Then, we start from an admissible initial configuration where no two robots are located on the same node and derive executions that never satisfy the exploration specification.\\

We consider the two similar configurations shown in figures \ref{fig1} and \ref{fig2}. As the configurations contain two axes of symmetry, the four robots $R1$, $R2$, $R3$ and $R4$ have identical views, which means that if they are activated simultaneously, they will exhibit the same behavior. 

\begin{figure}
 \begin{minipage}[b]{.46\linewidth}
  \centering\epsfig{figure=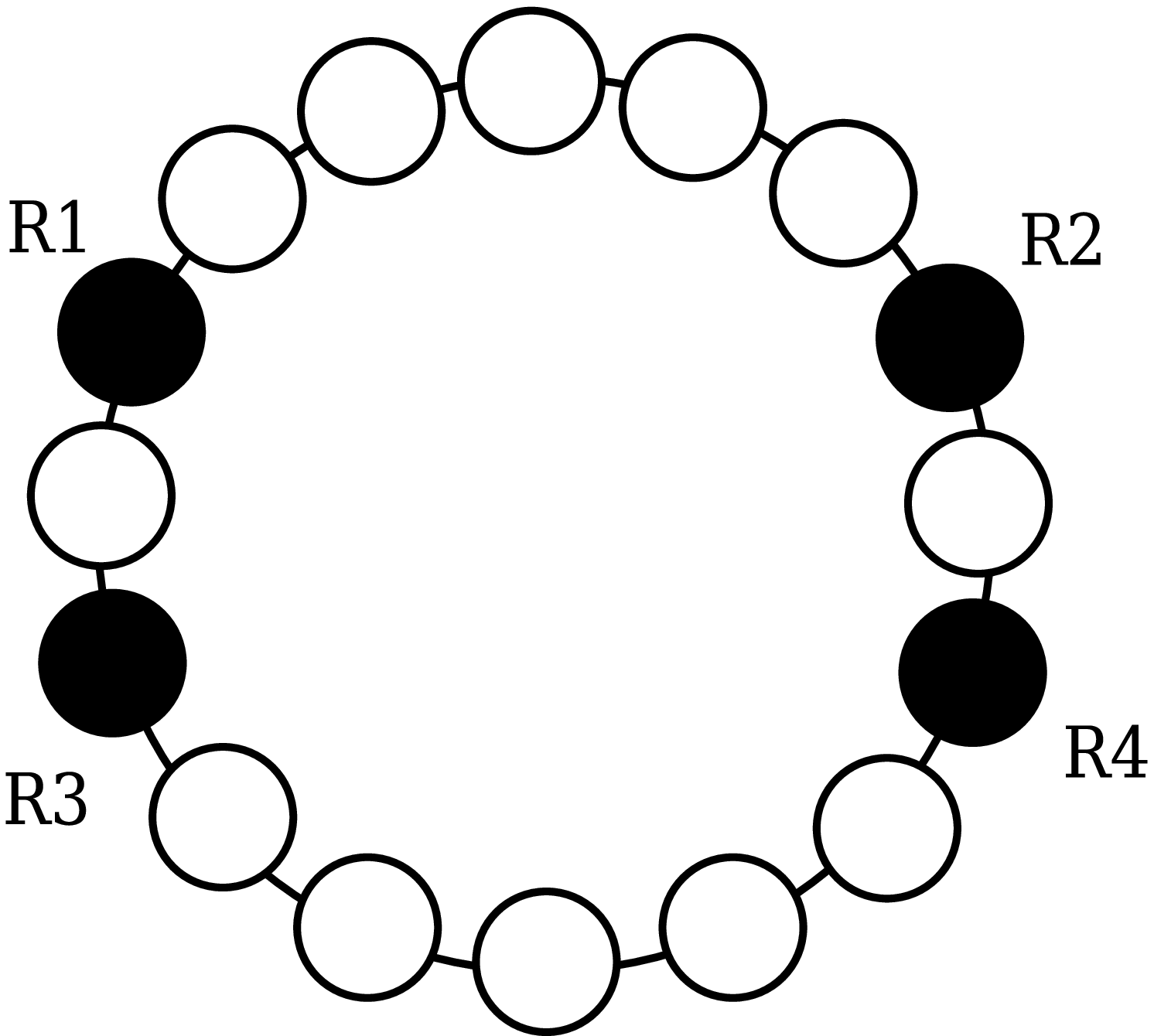,width=2.5cm}
  \caption{Instance of configuration \label{fig1}}
 \end{minipage} \hfill
 \begin{minipage}[b]{.46\linewidth}
  \centering\epsfig{figure=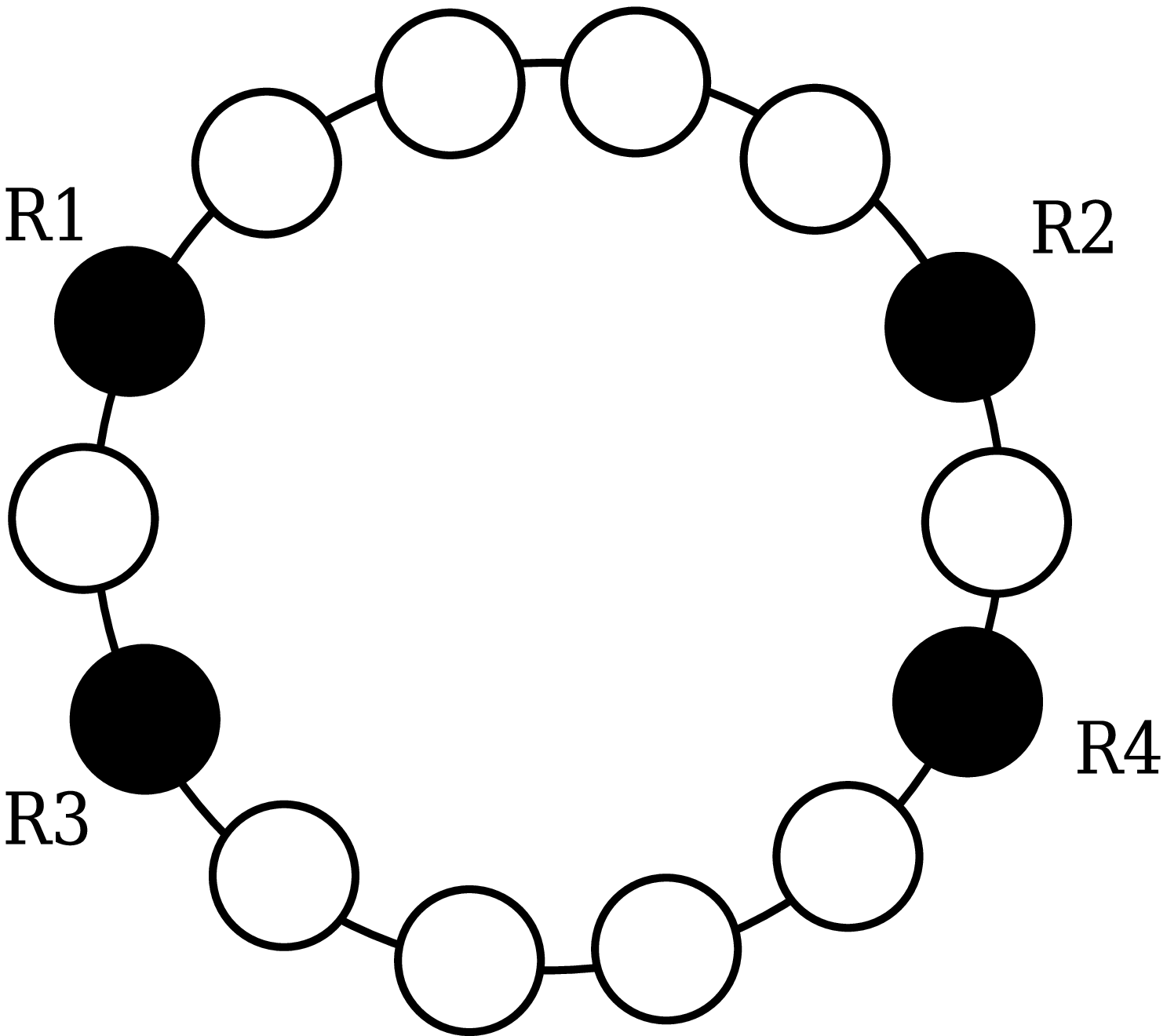,width=2.5cm}
  \caption{Instance of configuration  \label{fig2}}
 \end{minipage}
\end{figure}

\begin{enumerate}
\item {\textit{Suppose that every robot move towards it neighbor robot at distance $2$}}. Assume that all the robots are activated at the same time by the scheduler, then two towers are created (one with $R1$ and $R3$, the other with $R2$ and $R4$). From this point onwards, we assume that $R1$ and $R3$ are always activated simultaneously (and likewise for $R2$ and $R4$). As a result, the fours robots now behave as two robots. As it was shown in~\cite{stap09}, no team of two robots can explore the ring, and thus the initial protocol does not perform a ring exploration either.
\item {\textit{Suppose that every robot move towards their adjacent node in the opposite direction of their neighbor robot at distance $2$}}. If the robots move back to their position, then the protocol can never stop since the robots can go back and forth indefinitely. In the case where the robots keep moving away then two cases are possible : 
\begin{itemize}
\item {\textit{The number of nodes between $R1$ and $R2$ is even} (the same for $R3$ and $R4$ see figure \ref{fig2})}  in this case, by moving away from the robots that are at an odd distance from them, the configuration reached is similar to the one shown in figure \ref{fig:figr3} in which $R1$ and $R2$ are neighbors (the same for $R3$ and $R4$). Since the robots cannot go back (the protocol may never stop), the only move that they can perform is moving towards their neighbor: $R1$ moves towards $R2$ and vice versa (the same for $R3$ and $R4$), however, in the case where the four robots are activated at the same time, the two robots that are neighbors simply exchange their positions, and the configuration remains unchanged. As a result, no progress is made towards completion of the exploration task. 
  
\begin{figure}[H]
  \centering
  \includegraphics[width=2.5cm]{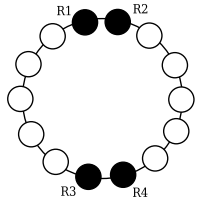}
    \caption{Instance of configuration} 
    \label{fig:figr3}
\end{figure}

\item {\textit{The number of nodes between $R1$ and $R2$ is odd} (the same for $R3$ and $R4$, see figure \ref{fig1})}. In this case $k$ divides $n$, and \cite{davi07} proved that the exploration problem in impossible to solve in this setting.
\end{itemize}  
\end{enumerate}

From the cases above, we can deduct that no  deterministic exploration is possible using four robots when the size of the ring is even.


\section{Ring Exploration in CORDA model}\label{sec:Exp}
In this section we propose the ring exploration in Corda model with only five robots. Before detailing our algorithm we introduce some definitions. A {\bf hole} is the maximal set of consecutive empty nodes, the size of a hole is the number of nodes that compose it, the border of the hole are the two empty nodes who are part of this hole, having one robot as a neighbor. An {\bf inter-distance} $d$ is the minimum distance taken among distances between each pair of distinct robots (in term of the number of edges). A {\bf d.block} is any maximal elementary path in which there is one robot every $d$ edges. The {\it border} of a d.block are the two robots belonging to this d.block having a hole of size bigger or equal to $d$ as a neighbor. A {\bf tower-chain} consists of an 1.block of size $3$ followed by an empty node followed by a tower.

\label{sec:Overview}
Our protocol consists of three distinct phases:
\begin{itemize}
\item{\textbf{Block Module}. The aim of this phase is to drag all the robots in one single $1$.block starting from any initial configuration that doesn't contain any tower.}

\item{\textbf{Tower Module}. Starting from a configuration that contains a single $1$.block, one tower is created in such way to give an orientation to the ring  allowing the elected robot to explore the ring in the last phase.}

\item{\textbf{Tower-chain Module} In this phase, starting from a configuration with a single tower, one robot is elected in order to explore the ring. 
}
\end{itemize}

{\footnotesize 
\begin{algorithm}[H]
  \caption{The orchestration of the algorithm}
  \label{protocol.algo}
 \begin{algorithmic}[1]
\If {the five robots do not form a tower-chain}
     \If {the configuration contains neither a tower nor a single $1$.block}
         \State \textbf{Execute} \textbf{Block Module}
      \Else
     \If {the configuration contains a single $1$.block}
          \State \textbf{Execute} \textbf{Tower Module} 
       \Else
           \State \textbf{Execute} \textbf{Tower-chain Module} 
     \EndIf
    \EndIf
   \EndIf
  \end{algorithmic}
\end{algorithm}
}
Note that once a configuration with a tower-chain is reached, the ring has been explored and the protocol terminates. Remark also that robots are able to distinguish the phase they are since each phase has different particularities. In the first phase all the configurations are tower-less and do not contain $1$.block of size $5$. 
In the second phase, configurations contain a single $1$.block of size $5$. And finally, in the last phase, the configurations contain a single tower.  

The following section details and analysis the complexity of the previous modules.

\paragraph{Block Module description and analysis.}\label{sec:ph1}

The aim of this phase is to reach a configuration where there is a single $1$.block that contains all the five robots without creating any tower. 
This phase is described in Algorithm~\ref{ph1.algo}.

{\footnotesize
\begin{algorithm}[H]
  \caption{Procedure: Block Module}
  \label{ph1.algo}
 \begin{algorithmic}[1]
\If {the configuration contains at least one isolated robot}
    \If {the configuration contains a single d.block}
        \If {I'm the isolated robot and I'm the closest neighbor to the $d$.block}
          \State \textbf{Move} toward the $d$.block taking the shortest hole
         \EndIf
    \Else
     \If {the configuration contains two $d$.block}
        \If {the configuration is symmetric}
            \If {I'm the isolated robot}
                \State \textbf{Move} toward one of the two $d$.blocks
             \EndIf
         \Else
             \State \textbf{Move} toward the closest $d$.block
          \EndIf
       \EndIf
     \EndIf
\Else 
    \If {the configuration contains a single $d$.block and $d>1$}
        \If {I'm at the border of the $d$.block}
            \State \textbf{Move} toward my adjacent node in the direction of the $d$.block
         \EndIf
    \Else
      \If {the configuration contains two $d$.blocks}
          \If {I'm the smallest $d$.block and the closest to the biggest $d$.block} 
              \State \textbf{Move} toward the biggest $d$.block   
          \EndIf
       \EndIf
     \EndIf
    \EndIf
  \end{algorithmic}
\end{algorithm} 
}
\begin{lem} If the configuration at instant $t$ contains neither a single $1$.block nor a tower, then the configuration at instant $t+1$ is tower-less.
\end{lem}

\textit{\textbf{proof:}} We prove in this section that, if a robot moves, it moves always to an empty node to avoid the creation of towers. We suppose that the configuration at instant $t$ is $C$. The configuration $C$ doesn't contain any tower and satisfies one of the following cases: 

\begin{itemize}
\item{$C$ contains at least one isolated robot:} Two cases are possible according to the number of $d$.blocks:
\begin{enumerate}
\item{There is a single $d$.block:} in this case the isolated robots that are the closest to the $d$.block are allowed to move, they move to an empty node (see line $4$). As it is an isolated robot, there are at least $d$ empty nodes between it and the target d.block. In another hand, since $d \geq 1$, by moving, no tower is created at instant $t+1$.
\item{There are two $d$.blocks:} in this case, the configuration contains a single isolated robot (there are five robots on the ring), this robot is the only one allowed to move (see line $9,10$), when it moves, it does to an empty node toward one of the two blocks depending of the symmetry of the configuration (there is no other robot between it and the target $d$.block and there are at least $d$ empty nodes between it and the $d$.blocks -- otherwise it would be part of them). Thus, no tower is created at instant $t+1$.
\end{enumerate}
\item{$C$ contains no isolated robots in the configuration:} two cases are possible according to the number of $d$.blocks:
\begin{enumerate}
\item{$C$ contains a single $d$.block:} in this case the robots at the border of this $d$.block are the only robots allowed to move, if they do, they move to an empty node toward the $d$.block they belong to. This guarantee is given by the condition $d>1$ (see line $18$).
Hence, no tower is created at instant $t+1$.
\item{$C$ contains two $d$.blocks:} in this case the two $d$.blocks have different sizes (since there are no isolated robots and the number of robots is odd). Robots in the smallest $d$.block and closest to the biggest $d$.block move toward the target $d$.block taking the hole that separate them from one extremity of the biggest $d$.block. Since the size of the hole is at least equal to the inter-distance (otherwise robots are in the same $d$.block), no tower is created at instant $t+1$.
\end{enumerate}
\end{itemize}
Overall no tower is created at instant $t+1$.

\begin{lem}
\label{lem:ph2}
Starting from a configuration with no tower the system reaches a configuration with a single $1$.block after $O(n)$ move operations.
\end{lem}

\textit{\textbf{proof:}} Two cases are possible according to the type of the starting configuration denoted in the following $C$:
\begin{enumerate}
\item{$C$ contains at least one isolated robot:} 
in this case, the robots allowed to move are always the isolated ones, and their destination is the closest $d$.block ($line 3,4$), or one of the two $d$.blocks in the case of symmetry ($line 9,10$). Hence three cases are possible according to the number of isolated robots:
\begin{itemize}
\item{The configuration $C$ contains a single isolated robot. This robot is the only one allowed to move and its destination is the $d$.block. After its move the distance between it and the target $d$.block decreases. Since this robot remains the only isolated robot in the configuration (robots in the $d$.blocks do not move when there is at least one isolated robot), it is the only one that keeps moving to the same target $d$.block. Therefore, after a finite time, the robot joins the $d$.block. 

In order to compute the maximum number of moves ($NBM$) we consider the 
the worst case: the number of nodes between the two robots at the border of the $d$.block is odd.  
Consider the Figure \ref{fig:case1}.

\begin{figure}[H]
  \centering
  \includegraphics[width=2.5cm]{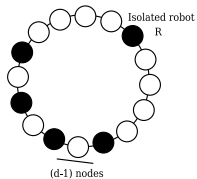}
    \caption{Instance of configuration} 
    \label{fig:case1}
\end{figure}

Let compute the number of nodes between the $d$.block and the isolated robot. In order to do this, we will substract from the size of the ring, the occupied nodes and the empty nodes between the robots in the $d$.block. 

The obtained result gives the sum of the empty nodes between the isolated robot and the $d$.block at each side. Thus, we have to divide it by two to obtain the distance between the isolated robot and the $d$.block. Note that the isolated robot is going to join the $d$.block hence it will advance until it reaches the same distance as the other robots in the d.block. Thus, in order to calculate the number of moves of the isolated robot, we have to substract from $n-[k+3*(d-1)]$, the number of nodes between any two robots in the $d$.block. 



The $NBM$ is hence given by the following formula: 
\begin{equation}
    NBM=[\frac {n-[k+3*(d-1)]}{2}]-(d-1)
  \label{max1}
 \end{equation}
}

\item{The configuration contains two isolated robots: in this case, at least one of these two isolated robots is allowed to move. If there is a single robot that is the closest to the $d$.block, then this robot is the only one that moves, its destination is the single $d$.block (see line $3,4$). At each move, the robot becomes even closer to the $d$.block, hence after a finite time, it reaches the $d$.block and the configuration contains a single isolated robot. In the other case (there are two robots allowed to move, let them be $R1$ and $R2$), whatever the choice of the scheduler in interleaving the different operations, at least one of these two isolated robots moves towards the $d$.block and hence becoming even closer. Note that if robots move at the same time both reduce their distance to the $d$.block. 

Suppose the worst case: a single robot moves. Let $R1$ be this robot. Two cases are possible: If the robot that does not move, $R2$, has an up to date  view of the configuration, then $R1$ becomes the closest robot to the d.block and hence it is the only one allowed to further move. From this point onward the proof is similar to the case 1. If $R2$ has an absolute view then $R2$ may also move. Consequently, either it is at the same distance as $R1$ from the $d$.block or $R1$ is the closest one to the $d$.block. The proof goes on recursively until at least one of the two robots reaches the $d$.block. 

Note that the worst case happens when the two robots are at the maximum distance from the $d$.block and the number of empty nodes between these two robots is minimal (equal to $d$ otherwise they form a $d$.block). It follows that the maximum number of moves robots perform in this case  is given by the following formula: 
\begin{equation}
   NBM= n-(k+4*(d-1)+d)
  \label{max2}
 \end{equation}
}
\item{The configuration contains three isolated robots. From the above cases above it follows that after a finite time all isolated robots join the $d$.block. The maximum number of moves in this case is performed when the three robots are at a maximum distance from the $d$.block and the distance between them is minimal and it is equal to $d$. This number is given by the following formula: 
\begin{equation}
    NBM= n-(k+3*(d-1)+2d)+[n-(k+3*(d-1))]/2-(d-1)
   \label{max3}
 \end{equation}
}
\end{itemize}
Overall the number of isolated robots decreases until the configuration contains only $d$.blocks.

\item{The configuration contains only $d$.blocks:} Two sub-cases are possible according to the number of $d$.blocks:
\begin{itemize}
\item{The configuration contains two $d$.blocks.} In this case the two $d$.blocks have different size and the smallest $d$.block moves towards the biggest one ($line 24,25$). Consequently, whatever the choice of the scheduler at least one of the two robots moves towards the biggest one, when it does the configuration changes and contains a single $d$.block with isolated robots. However, it has been shown in case $1$ that in this case and after a finite time all the isolated robots join the $d$.block. Hence the number of $d$.blocks decreases and the configuration contains a single $d$.block. The maximum number of moves is defined by the following formula and it happens when the small block is at a maximum distance from the biggest $d$.block: 
\begin{equation}
    NBM= n-(k+5*(d-1))
   \label{max4}
 \end{equation}
\item{The configuration contains a single $d$.block:} if $d>1$ then there is at least one single node between each robot, and in this case, depending on the choice of the scheduler at least one of the two robots that are at the border of the $d$.block moves to its adjacent node in the direction of the $d$.block it belongs to ($line 19,20$). Thus, the inter-distance decreases and the configuration reached contains isolated robots. However, once the configuration with a single $d$.block is reached, the maximum number of moves that are performed in order to reach another configuration with a single $(d-1)$.block $\forall d$ such as $d>1$ is constant and is equal to $7$. If one of the two robots at the border of the $d$.block moves, then one $(d-1)$.block is created and all the other robots move towards it starting with the closest one, since all the robots were at the same distance, the closest one performs one move to reach the $(d-1)$.block, the next robots performs two moves and the third one (the last one) performs three moves to reach the $d$.block. Hence, if we sum up all these moves taking in account the first move of the robot that creates the $(d-1)$.block, then the total number of displacements is the following: $1+(1+...+ k-2)$ and is equal to $7$. In the case where the two robots at the border of the d.block move at the same time, two $(d-1)$.block are created. The isolated robot that is on this axes of symmetry chooses one of them by moving towards it, when it moves it joins the chosen $(d-1)$.block and the configuration contains two $(d-1)$.blocks. However, in this case, only one robot is allowed to move (the closest one (see line $24,25$), since the two robots in the smallest $(d-1)$.block are at different distance from the biggest one). This robot performs two moves to join the biggest $(d-1)$.block, the same for the second robot. Thus if we sum up the moves that were performed ($2+1+2+2$), the total number is equal to $7$. Since $d>1$ the same process repeats the system reaches a configuration with a single $1$.block. 
In this case the number of moves is given by the following formula:
\begin{equation}
    NBM= (d-1)*7
   \label{max5}
 \end{equation} 
\end{itemize}
\end{enumerate}




Since $d< n/5-1$ the total number of moves in order to reach a $1.block$ configuration starting from any tower-less configuration is  $O(n)$.

\paragraph{Tower Module description and analysis.}\label{sec:ph2}

This phase begins when the configuration contains a single $1$.block. It aims at creating a tower in order to give a virtual orientation to the ring such as the elected robot accomplish the exploration task in the last phase. 
This phase is described in Algorithm \ref{ph2.algo}.
{\footnotesize
\begin{algorithm}[htbp]
  \caption{Procedure Tower Module}
  \label{ph2.algo}
 \begin{algorithmic}[1]
\If {I'm on the axes of symmetry}
    \State \textbf{Move} toward one of my neighbors
\EndIf
  \end{algorithmic}
\end{algorithm} 
}
\begin{lem}
\label{lem:tower}
Let $C$ be the configuration that contains a single 1.block of size $5$. If $C$ is the configuration at instant $t$, then the configuration at instant $t+1$ contains a single tower. 
\end{lem}

\paragraph{Tower-chain Module description and analysis.}\label{sec:ph3}
In this phase, one robot is elected in order to explore the ring. The exploration begins when the configuration contains a tower and is done when a tower-chain is created. This phase is described in Algorithm \ref{ph3.algo}.
{\footnotesize
\begin{algorithm}[htbp]
  \caption{ Procedure tower-chain Module}
  \label{ph3.algo}
 \begin{algorithmic}[1]
  \If {the configuration doesn't contain a chain-tower} 
     \If {I'm between the tower and the $1$.block}
        \State \textbf{Move} toward my adjacent node in the opposite direction of the tower 
       \EndIf 
      \EndIf
       \end{algorithmic}
\end{algorithm} 
}
\begin {lem}
\label{lem:ph3}
Starting from a configuration with a single tower, the system reaches a configuration that contains a chain-tower after $O(n)$ move operations and all the nodes have been explored.
\end {lem}


\end{proof}

\section{Conclusion}
In this paper, we focused on the exploration problem in an undirected ring. We proved that no deterministic protocol can explore such a graph using $k$ robots such as $k\leq 4$ if the ring is of even size. On the other hand, we provided a non-atomic completely asynchronous algorithm that uses only five robots for completing exploration provided that $n$ and $k$ are coprime. Our solution is thus optimal with respect to the number of robots. As exploration requires $O(n)$ robots moves, it is also optimal in time. We would like to mention two interesting open questions raised by our work:
\begin{enumerate}
\item The impossibility result of \cite{davi07} shows that $k$ must not divide $n$ (for arbitrary values of $k$ and $n$), while our impossibility result shows that $k$ must be coprime with $n$ (for a specific value of $k$: $4$). We conjecture that the impossibility result of \cite{davi07} can be extended to any $k$ and $n$ that are not coprime and such that $n>k$. 
\item Our impossibility result holds for even sized rings only, thus remains the open question of designing a deterministic exploration protocol with fours robots for odd sized rings in the CORDA model (note that this is feasible in the ATOM model as presented in the appendix).
\end{enumerate}

{\small
\bibliographystyle{plain}
\bibliography{biblio} 
}

\newpage

\appendix

\section{Proof of Lemma~\ref{lem:tower}}

let $R1$ be the robot located in the middle of the 1.block in $C$. In $C$ all robots execute algorithm $3$ (see algorithm $1$). Thus from $C$, $R1$ is the only one that can move, its destination is one of its adjacent node (see algorithm $3$), by moving, a tower is created (the adjacent node was occupied) and the lemma holds. 

\section{Proof of Lemma~\ref{lem:ph3}}

When a tower is created as output of the Tower Module, there will be a single robot between the tower and the $1$.block (let $R1$ be this robot). $R1$ is the only robot allowed to move, and if it does, it moves to its adjacent node in the opposite direction of the tower (see $line 2, 3$). After executing these actions, $R1$ remains the only robot between the tower and the $1$.block. Consequently, it keeps moving to the same target destination (the $1$.block). However, since at each move, the elected robot approaches the $1$.block and since there are $(n-5)$ empty nodes between $R1$ and the $1$.block at the beginning of tower-chain module, $R1$ will reach the $1$.block and a chain-tower will be created after $n-5$ movements. Also, since the $n-5$ nodes that were between the isolated robot and the $1$.block are the only nodes not explored (the five nodes that have been occupied in the beginning of phase $2$ are already explored), when a configuration with a chain-tower is reached, all nodes of the ring have been explored and the protocol terminates (see algorithm $4$ line $1$).

\newpage

\appendix

\section{Exploration-Odd in ATOM model}\label{sec:ExpOdd}

In this section we propose a \textbf{deterministic} algorithm for the ring exploration problem using only \textbf{four} robots in the ATOM model. We consider rings of 
odd size $n$ such that $n>7$ . 
 	
\begin{figure}
 \begin{minipage}[b]{.46\linewidth}
  \centering\epsfig{figure=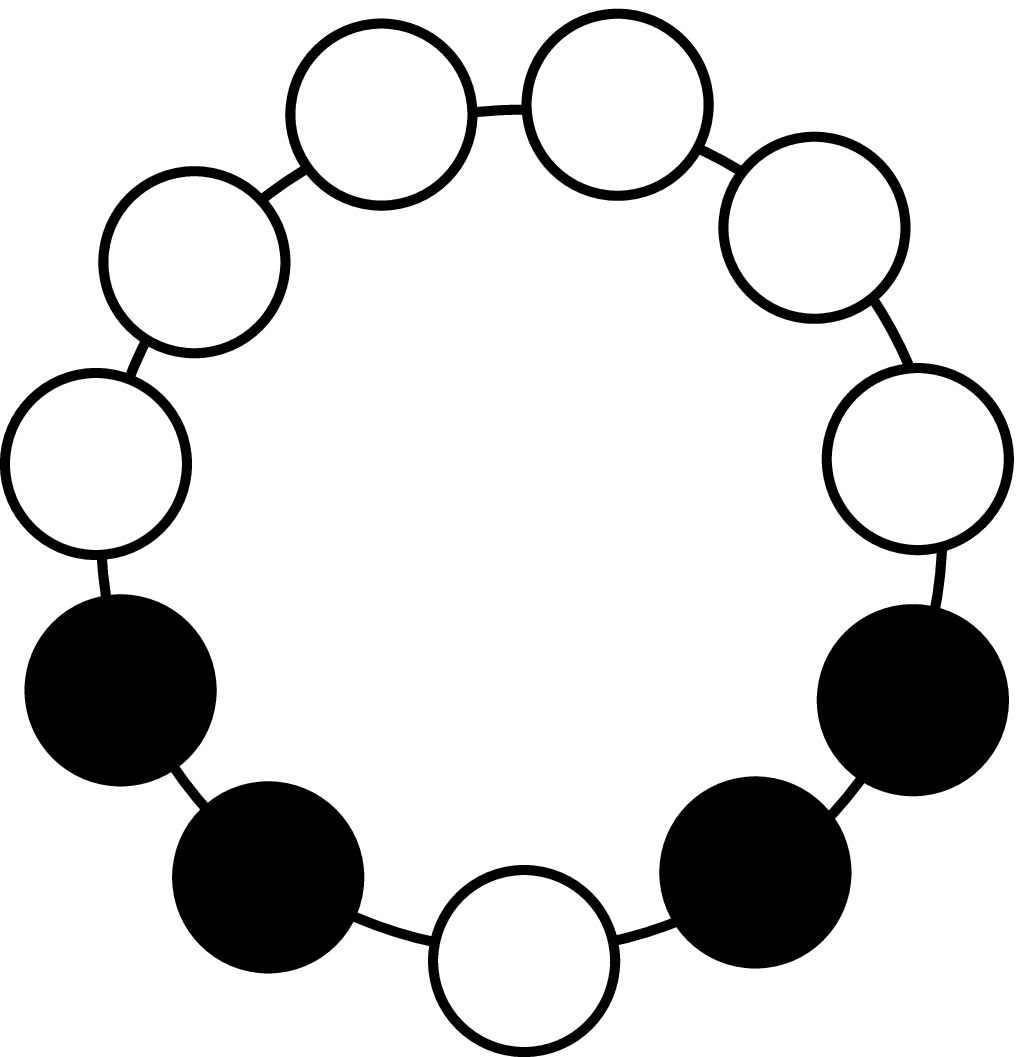,width=2cm}
  \caption{S-tower-plan\label{STP}}
 \end{minipage} \hfill
 \begin{minipage}[b]{.46\linewidth}
  \centering\epsfig{figure=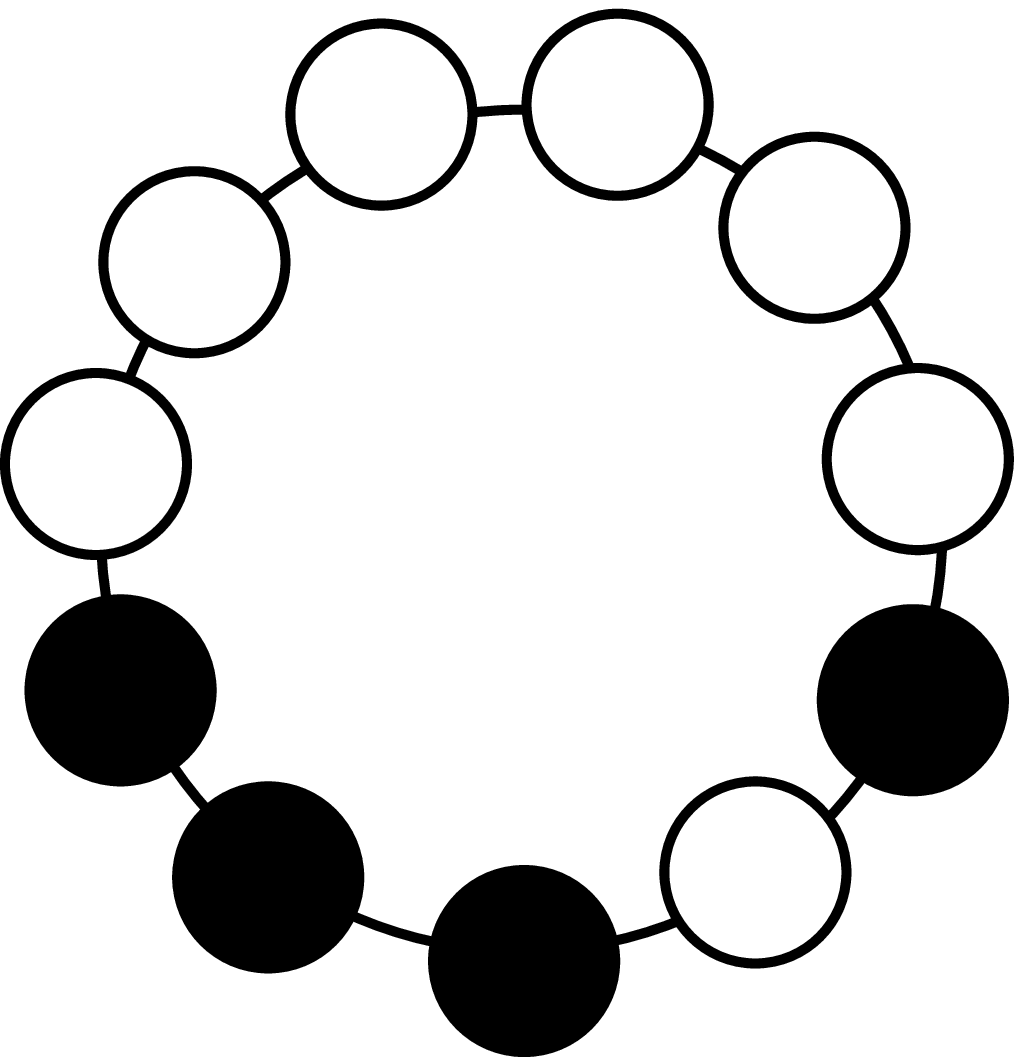,width=2cm}
  \caption{A-tower-plan  \label{ATP}}
 \end{minipage}
\begin{minipage}[b]{.46\linewidth}
  \centering\epsfig{figure=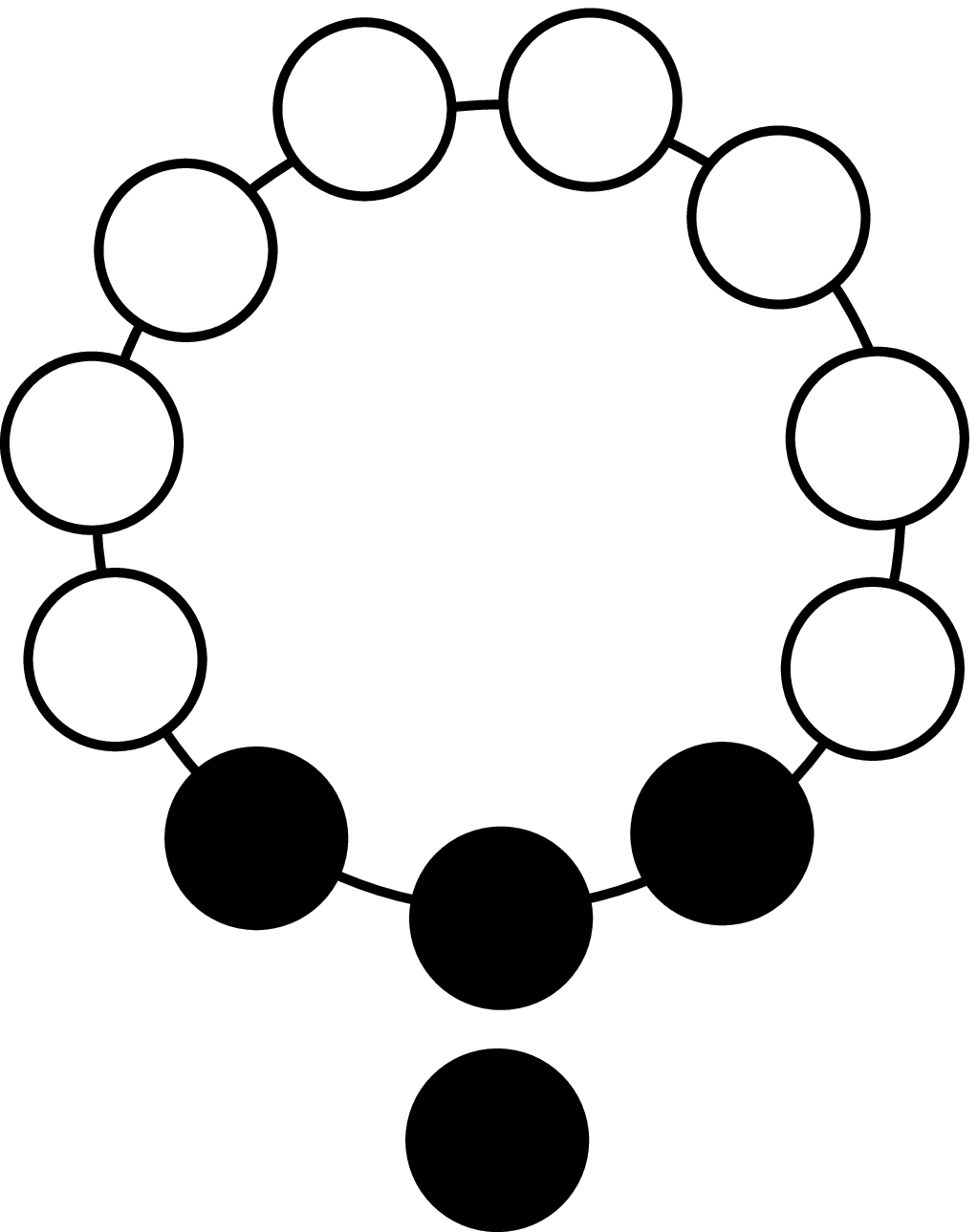,width=2cm}
  \caption{Tower-block\label{TB}}
 \end{minipage} \hfill
 \begin{minipage}[b]{.46\linewidth}
  \centering\epsfig{figure=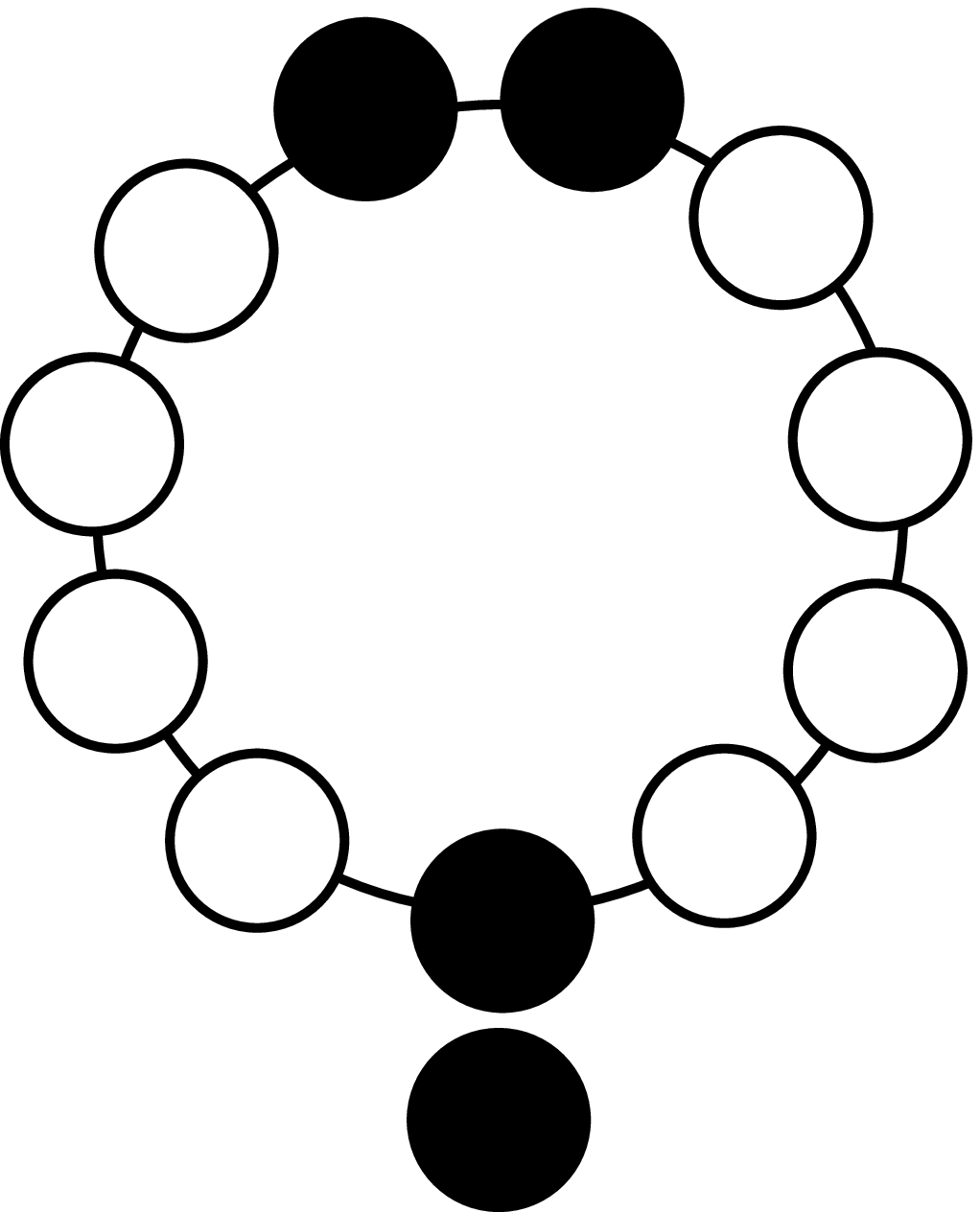,width=2cm}
  \caption{Tower-Sole  \label{TS}}
 \end{minipage}
\end{figure}

Before describing the algorithm, we first give some useful terms:
A configuration contains an {\bf S-tower-plan} if an only if the configuration is symmetric and the configuration contains two 1.blocks that share a hole of size $1$ (see Figure \ref{STP}). We say that a configuration contains an {\bf A-tower-plan} if and only if the configuration is not symmetric and contains one 1.block of size $3$ having an isolated robot as a neighbor at distance $2$ (see Figure \ref{ATP}). A {\bf tower-guide} is a path $u_{i},u_{i+1},u_{i+2},u_{i+3},u_{i+4}$, such as $u_{i+2}$ and $u_{i+4}$ are occupied by a single robot, $u_{i}$ contains a tower, $u_{i+1}$ and $u_{i+3}$ are empty nodes. A configuration contains a {\bf tower-block} if at each side of the tower there is one robot at distance $1$ (see Figure \ref{TB}). The configuration contains a {\bf tower-sole} if the configuration is symmetric and there is an 1.block that does not contain the tower (see Figure \ref{TS}).  

\subsection{\textbf{Overview of the algorithm}}
Exploration-odd protocol consists of three phases:
\begin{itemize}
\item{\textbf{Phase 1}}: The aim of this phase is to reach a configuration with a S-tower-plan or an A-tower-plan within a finite time starting from any initial towerless configuration.\\

\item{\textbf{Phase 2}}: Starting from a configuration that contains an (A or S)-tower-plan a tower is created. This tower is further used in the last phase.\\

\item{\textbf{Phase 3}}: In this phase the exploration is performed by one or two robots according to the location of the tower. The protocol terminates when the configuration contains a tower-block or a tower-sole. \\

\end{itemize}

\begin{algorithm}[H]
 \caption{The protocol}
  \label{GenAlgo.algo}
 \begin{algorithmic}[1]

\If {the configuration contains neither a tower-block nor a tower-sole}
    \If {the configuration contains neither a S-tower-plan nor an A-tower-plan}
         \State \textbf{Execute} Procedure Phase 1
     \Else
        \If {the configuration does not contain a tower}
         \State  \textbf{Execute} Procedure Phase 2
      \Else 
           \State {Execute} Procedure Phase 3
      \EndIf
\EndIf
\EndIf
\end{algorithmic}
\end{algorithm}

\subsection{\textbf{Detailed description of the solution}}

In this section, we describe the different phases of the algorithm with more details, we provide an algorithm for each phase and we prove the correctness of our protocol.
\subsubsection{\textbf{Phase1}}
The aim of this phase is to reach a configuration with a (S or A)-tower-plan without creating any tower during the process. This phase is described in Algorithm \ref{PH1.algo}. Let us 
define a couple of terms usefull in the sequel of the presentation. $Distance(X1,X2)$ is a function that returns the distance between the two robots $X1$ and $X2$. $distance(X)$ returns the distance between the robots satisfying property $X$. X=$RobotAxes$ refers to the two robots that are on the same side of the axes of symmetry, X=$SymRobotEv$ refers to the two symmetric robots that are at an even distance and X=$SymRobotOd$ refers to the two symmetric robots that are at an odd distance.\\

In the proofs below, in the case when the configuration is symmetric, we suppose that the two symmetric robots that are at an even distance from each other are $R1$ and $R2$, and the two other symmetric robots (that are at an odd distance) are $R3$ and $R4$. $R1$ and $R3$ are on the same side of the axes of symmetry (the same holds for $R2$ and $R4$). Recall that since the configuration is symmetric, $R1$ and $R2$ have identical views, which means that if they are activated simultaneously, they will exhibit the same behavior (the same for $R3$ and $R4$).

\begin{algorithm}[H]
  \caption{Procedure Phase 1}
  \label{PH1.algo}
 \begin{algorithmic}[1]
\If {the configuration is symmetric}
    \If {$distance(symrobotEv)= distance(symrobotOd)+ 1$}
         \If {$distance(robotAxes)$ is even}
             \If {$Distance(me,symRobot)$ is even}
                 \If {$Distance(me,symRobot)> 2$}
                         \State \textbf{\textit{Move}} towards my symmetric robot
                  \Else
                         \State \textbf{\textit{Move}} in the opposite direction of my symmetric robot
                  \EndIf
              \EndIf
          \Else
              \If {$distance(robotAxes)!=1$} 
                  \If {$Distance(me,symRobot)$ is odd}
                      \State \textbf{\textit{Move}} in the opposite direction of my symmetric robot
                  \EndIf
              \Else 
                   \If {$Distance(me,symRobot)$ is even} 
                       \State \textbf{\textit{Move}} towards my symmetric robot
                   \EndIf
              \EndIf
          \EndIf
      \Else
        \If {$Distance(me,symRobot)$ is even and is bigger than $2$}
            \If {I'm in an 1.block}
                \State \textbf{\textit{Move}} to my adjacent empty node towards my symmetric robot
            \EndIf
        \Else
             \If {I'm not part of an 1.block that contains one robot that is not my symmetric robot}
                 \State \textbf{\textit{Move}} in the opposite direction of my symmetric robot
             \EndIf
       \EndIf
    \EndIf
\Else /*the configuration is not symmetric*/
     \If {the configuration contains a single d.block of size 2}
         \If {I'm the closest isolated robot}
              \State \textbf{\textit{Move}} toward the d.block taking the shortest path
          \EndIf
     \Else
         \If {the configuration contains a single d.block of size 3}
             \If {$d>1$}
                 \If {I'm at the border of the d.block}
                      \State \textbf{\textit{Move}} in the direction of the d.block I belong to
                 \EndIf
              \Else
                 \If {I'm the isolated robot}
                          \If {there is at least two nodes between me and the d.block on the shortest path}
                              \State \textbf{\textit{Move}} towards the 1.block
                          \EndIf
                     \EndIf
                \EndIf
\EndIf
\EndIf
\EndIf
\end{algorithmic}
\end{algorithm}

\begin{lem}
If the configuration at instant $t$ contains neither a tower nor a (S or A)-plan-tower then the configuration at instant $t+1$ is tower-less.
\end{lem}

\begin{proof}
Two cases are possible according to the type of the configuration

\begin{enumerate}
\item{\textit{the configuration is symmetric:}}  Two cases are possible:
\begin{itemize}
\item {$Distance(R1,R2)= Distance(R3,R4)+1$}
\begin{itemize}
\item{$Distance(RobotAxes)$ is even}. In this case, $R1$ and $R2$ are the robots allowed to move. If $Distance(R1,R2)>2$ then $R1$ and $R2$ move towards each other (see line $5,6$), hence no tower is created (there are at least three empty nodes between them since $Distance(R1,R2)>2$ and $R1$, $R2$ are at an odd distance). If $Distance(R1,R2)=2$, as there are four robots and the size of the ring is bigger than $7$, there are at least two empty nodes between the robots at each side of the axes of symmetry. Therefore no tower is created when $R1$ and $R2$ move towards $R3$ and $R4$ respectively (see line $8$). 

\item{$Distance(RobotAxes)$ is odd and is different from $1$}. In this case $R3$ and $R4$ are the two symmetric robots allowed to move. When they move, they go in the opposite direction of their symmetric robot (see line $13,14$). Since the distance between the robots at the same side of the axes of symmetry ($R1$ and $R3$, the same for $R4$ and $R2$) is different from $1$, then there is at least one empty node between each pair of robot at the same side of the axes of symmetry. Therefore, after the move is completed no tower is created.

\item{$Distance(RobotAxes)$}. As there are four robots and the size of the ring is bigger than $7$, there are at least three empty nodes between $R1$ and $R2$. Therefore, when robots move towards each other (see line $18,19$) no tower is created.
\end{itemize}

\item {$Distance(R1,R2)$ is different from $Distance(R3,R4)+1$}. When $R1$ and $R2$ move (they are part of an 1.block), then there are at least $3$ empty nodes between them (see line $23$), thus,by moving no tower is created. When $R3$ and $R4$ move, we are sure that they are not in an 1.block that contains $R1$ and $R2$ respectively, thus there is at least one empty node between them and the robot at the same side of the axes of symmetry. Therefore,  when robots move in the opposite direction of each other (see line $28,29$) no tower is created.

\end{itemize}
\item{\textit{the configuration is not symmetric}:} three cases are possible according to the type of the sub -configuration:
\begin{itemize}
\item{\textit{The configuration contains a single d.block of size $2$}}. In this case the closest isolated robot is allowed to move (there is only one since the configuration is not symmetric), when it does, it moves towards the d.block (see line $35,36$). However, since it is an isolated robot, we are sure that there are at least $d$ empty nodes between it and the d.block. On the other hand, since $d\geq 1$, then there is at least one empty node between the robot allowed to move and the target destination. Thus no tower is created.
\item{\textit{The configuration contains a single d.block of size $3$}}. In this case, if $d>1$, then the robots that are at the border of the d.block move in the direction of the block they belong to (see line $40,41,42$) and hence no tower is created (there is at least one empty node between robots at the border and the internal robot in the d.block since $d>1$). In the case where $d=1$, the isolated robot is the one allowed to move if and only if it is not at distance $2$ from the 1.block (see line $45,46,47$), hence there are at least two empty nodes between it and the 1.block. Thus when it moves, no tower is created.  
\end{itemize}
\end{enumerate}

Overall the configuration reached at instant $t+1$ contains no tower.
\end{proof} 

\begin{lem}
\label{lem:nosym}
Starting from a configuration that does not contain an axes of symmetry, a configuration with an A-tower-plan is reached in a finite time. 
\end{lem}

\begin{proof}
According to the size of the d.block in the configuration, two cases are possible:
\begin{enumerate} 
\item{\textit{the configuration contains a single d.block of size $2$}}. Since the configuration is not symmetric, there is exactly one robot which is the closest to the d.block. This robot is the only one allowed to move, its destination is the d.block (see line $(35,36)$). By moving, the distance between it and the target d.block decreases. Hence after a finite time, it joins the d.block and the configuration contains one d.block of size $3$ (we retrieve case ~\ref{block3}).
\item\label{block3}{\textit{the configuration contains a single d.block of size $3$}}. If $d>1$, then the robots that are at the border of the d.block move towards it (see line $(40,41,42)$). If the two robots are activated in the same time, then the configuration remains the same with a single (d-1).block of size $3$. In the case where only one robot is activated, then a new d.block of size $2$ is created with an inter-distance equal to $d-1$. In another hand, since the robot that was supposed to move is the closest one to the d.block, it is the only one allowed to move. After it completes the move it joins the new d-1.block, thus the configuration contains a single block of size $3$ with an inter-distance equal to $d-1$. Therefore after a finite time the configuration contains a single 1.block of size $3$ and one isolated robot. The isolated robot is the only robot allowed to move while it not at distance $2$ from the d.block (see line $(45,46,47)$). However, since at each move it becomes closer to the d.block, after a finite time, the isolated robot becomes neighbor of the 1.block at distance $2$ and thus the configuration contains an A-plan-tower and the lemma holds.
\end{enumerate} 
\end{proof}
\begin{lem}
\label{lem:EvsOd}
Starting from a symmetric configuration in which $distance(SymRobotEv)<distance(SymRobotOd)$, either the symmetry is broken or a configuration with an S-tower-plan is reached after a finite time. 
\end{lem}

\begin{proof}
Two cases are possible according to the distance between the robots that are at the same side of the axes of symmetry. 
\begin{itemize}
\item{$Distance(RobotAxes)!=1$. $R3$ and $R4$ are the only robots allowed to move, their destination is their adjacent node towards the robot being at the same side of the axes of symmetry. In the case where the scheduler activates only one of these two robots then the symmetry is broken and the lemma holds. Otherwise, the configuration remains symmetric and $Distance(R3,R4)>Distance(R1,R2)$. However, $distance(RobotAxes)$ decreases. Thus after a finite time, either the symmetry is broken or $Distance(R1,R3)=Distance(R2,R4)=1$.}
\item{$Distance(RobotAxes)=1$. In the case where there is one empty node between the robots that are at an even distance ($R1$ and $R2$), then the configuration is terminal (S-tower-plan is created) and the lemma holds. Otherwise, $R1$ and $R2$ are the only robots allowed to move, their destination is their adjacent node towards each other. If the scheduler activates one of them then the symmetry is broken and the lemma holds. Otherwise, the configuration remains symmetric, $Distance(R1,R2)<Distance(R3,R4)$ and $Distance(R1,R3)!=1$. In addition $Distance(R1,R2)$ decreases.}
\end{itemize}
Thus, in the case where the distance between the robots at the same side of the axes of symmetry is bigger than $1$, the robots that are at an odd distance are the ones that moves, by moving distance(RobotAxes) decreases until it becomes equal to $1$ (we suppose that the scheduler activates the symmetric robots at the same time, the symmetry is broken in the other case). When it is the case, the robots that are at an even distance are the one that move, by moving the distance between them decreases. Thus, starting from a symmetric configuration where $distance(RobotEv)<distance(RobotOd)$, either the symmetry is broken (in the case where the scheduler activates only one robot) or a terminal configuration is reached (configuration with an S-tower-plan). 
\end{proof}
\begin{lem}
\label{lem:EvOd1}
Starting from a symmetric configuration in which $distance(SymRobotEv)=distance(SymRobotOd)+1$, either the symmetry is broken or a configuration in which $distance(SymRobotEv)<distance(SymRobotOd)$ is reached in finite time. 
\end{lem}

\begin{proof}
Three cases are possible:
\begin{enumerate}
\item\label{even}{$Distance(RobotAxes)$ is even}. If $Distance(R1,R2)>2$ then $R1$ and $R2$ are the robots allowed to move towards each other (see line $5,6$). In the case when one of them is activated by the scheduler (suppose that $R2$ is the one that moves), a new axes of symmetry is created, $R1$ and $R3$ become symmetric robots (the same for $R4$ and $R2$). In the new configuration the distance between $R4$ and $R2$ is bigger than the distance between $R1$ and $R3$ and the lemma holds. In the case, when both  $R1$ and $R2$ are activated then by moving $Distance(R1,R2)< Distance(R3,R4)$ and the lemma holds. 

If $Distance(R1,R2)=2$, then $R1$ and $R2$ remain the robots allowed to move. However, their destination changes (they move in the opposite direction of each other see line $8$). In the case when the scheduler activated one of them the symmetry is broken and the lemma holds. Otherwise, $Distance(R1,R2)=Distance(R3,R4)+3$. Since $Distance(R1,R2)>Distance(R3,R4)+1$, $R3$ and $R4$ are the only one allowed to move. If the scheduler activates only one of them then the symmetry is broken and the lemma holds. Otherwise, the new configuration verifies $Distance(R1,R2)=Distance(R3,R4)+1$. However $Distance(R1,R2)>2$, thus either the symmetry is broken or $Distance(R1,R2)< Distance(R3,R4)$. In both cases the lemma holds.    

\item\label{Odd}{$Distance(RobotAxes)$ is odd and is different from $1$}. in this case the robots allowed to move are $R3$ and $R4$, their destination is the robot that is at the same side of the axes of symmetry ($R1$ and $R2$ respectively).In the case when they are activated at the same time, the configurations remains symmetric and $Distance(R3,R4) > Distance(R1,R2)$ (the lemma holds). If the scheduler activates only one of them (let this robot be $R4$) then a new axes of symmetry is created however the distance between the robots that are at odd distance ($R3$ and $R1$) is bigger than the distance between the robots at an even distance ($R2$ and $R4$), hence the lemma holds. 

\item\label{Odd1}{$Distance(RobotAxes)= 1$}. In this case the two robots $R1$ and $R2$ are the ones allowed to move (see line $17$,$18$), their destination is their adjacent node in the direction of each other. In the case where the scheduler activates both of them at the same time then the configuration remains symmetric and the distance between them becomes smaller than $Distance(R3,R4)$. Thus the lemma holds. In the case where the scheduler activates only one of them then, a new axe of symmetry is created and distance(robotAxes) becomes odd. However, it has been shown in ~\ref{Odd} that in this case either the symmetry is broken or $distance(SymRobotEv)<distance(SymRobotOd)$ and the lemma holds. 
\end{enumerate}
\end{proof}
\begin{lem}
\label{lem:EvbOd}
Starting from a symmetric configuration in which $distance(SymRobotEv)>distance(SymRobotOd)+1$, then after a finite time, either the symmetry is broken or $distance(SymRobotEv)=distance(SymRobotOd)+1$.  
\end{lem}

\begin{proof} Two cases are possible according to the distance of the robots at the same side of the axes of symmetry.
\begin{itemize}
\item{$distance(RobotAxes)!=1$. In this case, the two symmetric robots that are at an odd distance are the one allowed to move, their destination is their adjacent node in the direction of the robot that is at the same side of the axes of symmetry(see line $28$,$29$). If the scheduler activates only one of them, then the symmetry is broken and the lemma holds. Otherwise, either the configuration remains the same however $distance(RobotAxes)$ decreases. or $Distance(R1,R2)= Distance(R3,R4)+1$ or $Distance(R1,R2)>Distance(R3,R4)+1$ and $distance(RobotAxes)=1$. Hence after a finite time either the symmetry is broken or a configuration in which one of the two latter case is reached ($Distance(R1,R2)= Distance(R3,R4)+1$ (the lemma holds) or $Distance(R1,R2)>Distance(R3,R4)+1$ and $distance(RobotAxes)=1$).} 
\item{$distance(RobotAxes)=1$. $R1$ and $R2$ are the only robots allowed to move and they move towards each other. If the scheduler activates only one of them then the symmetry is broken and the lemma holds. In the other case, in the configuration reached either $Distance(R1,R2)= Distance(R3,R4)+1$ and we are done or $Distance(R1,R2)>Distance(R3,R4)$ and $distance(RobotAxes)!=1$. However the difference between distances that are between each pair of symmetric robots decreases.}  
\end{itemize}
From the two sub-cases above, we deduct that starting from a configuration in which $distance(SymRobotEv)>distance(SymRobotOd)$, either the symmetry is broken or a configuration in which $distance(SymRobotEv)= distance(SymRobotOd)+1$ and the lemma holds.  
\end{proof}


\begin{lem}
\label{lem:PH1}
Starting from any initial configuration that does not contain any tower, the system reaches a configuration with a (S or A)-plan-tower after a finite time.
\end{lem}

\begin{proof}
From lemma \ref{lem:nosym}, \ref{lem:EvsOd}, \ref{lem:EvOd1} and \ref{lem:EvbOd} imply that starting from any initial configuration that does not contain any tower, the system reaches a configuration with a (S or A)-plan-tower after a finite time.  
\end{proof}

\subsubsection{\textbf{Phase 2}}

This phase starts when the configuration contains a (S or A)-plan-tower. Its aim is to create a single tower in order to allow the exploration in the last phase.
In the case where the configuration is symmetric the two robots that share a hole of size $1$ move towards each other. If the configuration is not symmetric then the robot that is in the middle of 1.block moves towards its adjacent node not having a neighbor at distance $2$. This phase is described in algorithm \ref{PH2.algo}

\begin{algorithm}[H]
\caption{Procedure Phase 2}
\label{PH2.algo}
\begin{algorithmic}[1]

    \If {the configuration contains two 1.blocks}
        \If {I share a hole of size $1$ with my symmetric robot}
            \State \textit{\textbf{Move}} towards my symmetric robot
         \EndIf
     \Else
    \If {the configuration contains a single 1.block of size $3$}
        \If {I'm at in the middle of the 1.block}
             \State \textit{\textbf{Move}} towards my adjacent node that doesn't have a neighbor at distance $2$
        \EndIf
    \EndIf
\EndIf
\end{algorithmic}
\end{algorithm}

\begin{lem}
\label{lem:PH2}
Starting from a configuration that contains a plan-tower, the system reaches a configuration with a single tower after a finite time.
\end{lem}

\begin{proof}
Two cases are possible according to the type of the plan-tower:
\begin{enumerate}
\item{The configuration contains two 1.blocks}. The robots allowed to move in such a configuration are the ones that share a hole of size $1$ (see line $2,3$). In the case when they are activated at the same time by the scheduler, a single tower will be created on the axes of symmetry and the lemma holds. In the other case, the configuration reached contains one 1.block of size $3$ having an isolated robot as a neighbor at distance $2$, and we retrieve case ~\ref{APTower} 
\item\label{APTower}{The configuration contains a single 1.block of size $3$}. In this case the robot that is in the middle of the 1.block moves towards its adjacent occupied node that has not a neighbor at distance $2$. Hence a single tower is created and the lemma holds.
\end{enumerate}
\end{proof}
\subsubsection{\textbf{Phase 3}}

This phase is described in algorithm \ref{PH3.algo}. In this phase the ring is explored using one or two robots according to location of the tower.

\begin{algorithm}[H]
\caption{Procedure Phase 3}
\label{PH3.algo}
\begin{algorithmic}[1]
\If {there is no robot at distance $1$ from the tower}
    \If {the configuration contains a tower-guide}
        \If {I'm not at distance $2$ from the tower}
             \If {there is at least one empty node between me and the other isolated robot}
                 \State Move towards the isolated robot 
             \EndIf
        \EndIf
    \ElsIf {I'm at distance $2$ from the tower}
              \If {I'm in an 1.block}
                   \State Move towards the tower
               \Else
                    \State Move in the opposite direction of the tower
               \EndIf
    \Else
          \If {There is no robot at distance $2$ from the tower}
               \If {I'm an isolated robot and I'm not at distance $(n-1)/2$ from the tower} 
                    \State \textbf{\textit{Move}} to my adjacent node in the opposite direction of the tower
                \EndIf
           \EndIf
    \EndIf
\Else /*there is one robot at distance $1$ from the tower*/
     \If {I'm not at distance $1$ from the tower}
         \State Move towards the tower through the hole that we share
     \EndIf
\EndIf

\end{algorithmic}
\end{algorithm}   

\begin{lem}
\label{lem:PH3}
Starting from a configuration with a tower, the system reaches a configuration that contains a tower-block or a tower-sole after a finite time and all the nodes have been explored. 
\end{lem}  

\begin{proof}
At the beginning of the phase, we are sure that at least one isolated robot is at distance $2$ from the tower. If there is a single robot at distance $2$ from the tower (let this robot be R1), then the tower was created from an A-tower-plan and the configuration contains a tower-guide. Suppose that the other isolated robot is $R2$. In such a configuration R2 is the only one that can move, when it is activated it moves towards $R1$ (see line $5,6$) and a new 1.block is created. $R1$ is now, the only robot allowed to move, its destination is the tower (see line $10,11$). When $R1$ becomes neighbor of the tower, we are sure that the nodes between $R1$ and $R2$ have been explored. $R2$ becomes the only robot that can move and this while there is a hole between it and the tower, its destination is the tower, hence , when it moves, it becomes a direct neighbor of the tower (at distance 1 from the tower). Since there was $n-4$ nodes between the tower and $R2$ (all the nodes except the one that contains the tower, the nodes that are occupied by $R1$ and $R2$ and the empty node that was between $R1$ and $R2$ before moving $R2$), when $R2$ becomes neighbor of the tower, the configuration contains a tower-block and all the nodes have been explored.
In the case where there are two robots at distance $2$ from the tower (let them be $R1$ and $R2$), then the tower was created from a S-tower-plan and the two robots $R1$ and $R2$ are the only one allowed to move. Their destination is their adjacent node in the opposite direction of the tower (see line $12$). In the case where the scheduler activates only one of them, then the robot that is still at distance $2$ from the tower is the only one that can move keeping the same destination (see line $13$). After its move there will be no robots at distance $2$ from the tower (the same case occurs when the two robots are activated at the same time by the scheduler). The two isolated robots are allowed to move until they become at distance $(n-1)/2$ from the tower and the configuration will contain a tower-sole. Thus, every robots explored $(n-5)/2$ nodes, \textit{ie}, all the nodes have been explored. 
\end{proof}

\begin {lem}
Starting from any initial configuration, Odd-Exploration protocol terminates after a finite time.
\end {lem}
\textit{\textbf{proof:}}
Lemmas \ref{lem:PH1}, \ref{lem:PH2}  and \ref{lem:PH3} imply that, starting from any initial configuration that does not contain a tower, the system reaches a configuration with a tower-block or a tower-sole after a finite time. Note that all configurations with a tower sole or tower-block  are terminal (see algorithm \ref{GenAlgo.algo} line $1$). Thus, Odd-Exploration protocol terminates after a finite time.


\end{document}